\documentclass[a4paper,prl,superscriptaddress,showpacs,twocolumn]{revtex4}
\usepackage[dvips]{graphicx}
\usepackage{amssymb}
\newcommand{\ped}[1]{\ensuremath{_{\rm #1}}}

\begin{document}

\title{Evidence for two-gap nodeless superconductivity in SmFeAsO$_{0.8}$F$_{0.2}$
from point-contact Andreev-reflection spectroscopy}

\author{D.Daghero \email{E-mail:dario.daghero@polito.it}}
\author{M.Tortello}
\author{R.S. Gonnelli}

\affiliation{Dipartimento di Fisica and CNISM, Politecnico di
Torino, 10129 Torino, Italy}
\author{V.A.Stepanov}
\affiliation{P.N. Lebedev Physical Institute, Russian Academy of
Sciences, 119991 Moscow, Russia}
\author{N.D.Zhigadlo}
\author{J.Karpinski}
\affiliation{Laboratory for Solid State Physics, ETHZ, CH-8093
Zurich, Switzerland}

\begin{abstract}
Point-contact Andreev-reflection measurements were performed in
SmFeAsO$_{0.8}$F$_{0.2}$ polycrystals with T$_c$ $\simeq 53$ K. The
experimental conductance curves reproducibly exhibit peaks around
$\pm$ $6$ mV and shoulders at V $\sim 16-20 $ mV, indicating the
presence of two nodeless superconducting gaps. While the single-band
Blonder-Tinkham-Klapwijk fit can only reproduce a small central
portion of the conductance curve, the two-gap one accounts
remarkably well for the shape of the whole experimental d$I$/d$V$.
The fit of the normalized curves give $\Delta_1(0)=6.15\pm0.45$ meV
and $\Delta_2(0)=18\pm3$ meV. Both gaps close at the same
temperature and follow a BCS-like behavior.

\end{abstract}
 \maketitle

%\section{Introduction}

The experimental evidence of superconductivity in F-doped LaFeAsO
\cite{Kamihara_La} opened the way to the discovery of a new class of
superconductors that, with the exception of the copper-based
high-T\ped c superconductors, show the highest superconducting
critical temperatures known so far, with a record T$_c$ of 55 K in
F-doped or oxygen deficient SmFeAsO \cite{Ren_Sm,Ren_OxDef}.

Similarly to LaFeAsO \cite{de la Cruz}, SmFeAsO is semimetallic and
shows a spin-density-wave (SDW) antiferromagnetic order as well as a
tetragonal-to-orthorombic structural transition at $\sim 140 $ K
\cite{Drew_PhDiagr_Sm}. Above a critical value of F substitution
\cite{Drew_PhDiagr_Sm}, charge doping on FeAs planes suppresses the
SDW order and superconductivity occurs. The vicinity of the
superconducting state to a magnetic order raises many questions
concerning the pairing mechanism that is responsible for
superconductivity. Theoretical calculations \cite{Boeri_nophonons}
for the LaFeAsOF system showed that the electron-phonon coupling is
not sufficient to explain the observed T$_c$. An extended $s
\pm$-wave pairing with a sign reversal of the order parameter
between different sheets of the Fermi surface was thus proposed
\cite{Mazin_spm} for the same compound and shown to be compatible
with a coupling mechanism related to spin fluctuations. On the other
hand, a growing evidence for multi-gap superconductivity in
Fe-As-based compounds is being provided by several experimental
works
\cite{Hunte,Ding_ARPES_BaKFeAs,Szabo_BaKFeAs,LaOFFeAS,Kawasaki_NQR_LaOFFeAS,Matano_Pr_NMR,Samuely_Nd_PCAR,Pan_Nd_STM,Wang_PCS}.

In this situation, where there is no general consensus concerning
the pairing mechanism in these new superconductors, the
determination of the number, nature and symmetry of the
superconducting order parameter(s) assumes particular importance. In
this regard, point-contact Andreev-reflection (PCAR) spectroscopy is
a powerful technique to determine the gap value(s) and its (their)
symmetry. PCAR measurements performed so far in SmFeAsO$_{1-x}$F$_x$
\cite{Sm_Nature,Wang_PCS} are somehow in disagreement with each
other: Chen \emph{et al.} \cite{Sm_Nature} observed a single,
BCS-like, s-wave gap while Wang \emph{et al.} \cite{Wang_PCS}
reported two nodal order parameters. Here we report results of
point-contact Andreev-reflection measurements on high-quality
polycrystalline samples of SmFeAsO$_{0.8}$F$_{0.2}$ with T$_{c}^{on}
= 53$ K reproducibly showing the presence of two nodeless energy
gaps, $\Delta_1(0)= 6.15 \pm 0.45$ meV and $\Delta_2(0)=18 \pm 3$
meV; both gaps show a BCS-like temperature dependence and close at
the same temperature, always very close to the bulk T$_c$ measured
by susceptibility and resistivity.

\begin{figure}[t]
\begin{center}
\includegraphics[keepaspectratio, width=1 \columnwidth]{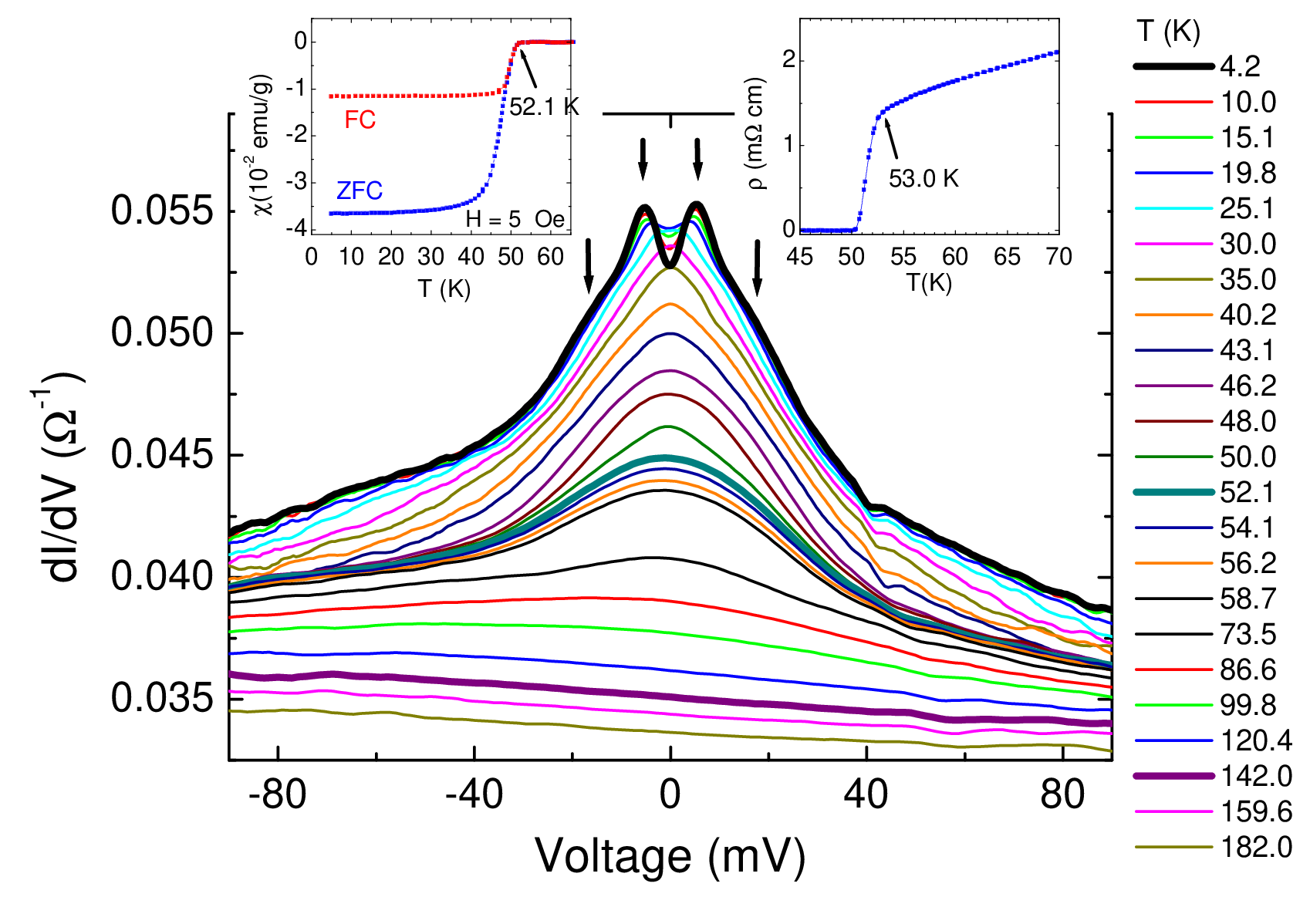}
\end{center}
\vspace{-5mm} \caption{\small{(color online) Temperature dependence
of the raw conductance curves of a point contact with T$_c \sim 52$
K. The low-temperature curve (upper thick line) features two peaks
and two shoulders (indicated by arrows) related to $\Delta_1$ and
$\Delta_2$, respectively. The curve at T$_c$ (middle thick line) has
a hump-like shape which flattens at $\sim 140$ K (lower thick line).
Insets show the resistivity and dc susceptibility of the sample.}}
\label{Fig1} %\vspace{-5mm}
\end{figure}

%\section{Experimental}
The polycrystalline samples of SmFeAsO$_{0.8}$F$_{0.2}$ were
synthesized under high pressure starting from SmAs, FeAs, SmF$_3$,
Fe$_2$O$_3$ and Fe. After the materials were pulverized and sealed
in a BN crucible, a pressure of 30 kbar was applied at room
temperature. The temperature was then increased within 1 h up to
1350-1450 $^{\circ}$C, kept for 4.5 h and quenched back to room
temperature. Finally the pressure was released. The resulting
samples are very compact and made up of shiny crystallites whose
size, as revealed by SEM images, is of the order of 30 $\mu$m. The
onset of the magnetic (resistive) transition is 52.1 K (53.0 K), as
shown in the insets of Fig. \ref{Fig1}.

Instead of a metallic tip pressed against the sample as in the
standard PCAR technique, we made the point contacts by placing a
small drop of Ag conducting paste on the freshly cleaved surface of
the sample. This is a pressure-less technique we successfully used
in various classes of superconductors, i.e. MgB$_2$, CaC$_6$,
ruthenocuprates, A15 (for details see, for instance, refs.
\cite{nostroPRL} and \cite{CaC6}). It gives much more stable
contacts under thermal cycling, which allowed us to easily record
the conductance curves up to 200 K.

When performing PCAR experiments, the size of the contact should be
smaller than the electronic mean free path, $\ell$, in the
superconductor \cite{Duif}. Only in this way the condition for
ballistic conduction in the contact is fulfilled and spectroscopy is
possible. Although at present the value of $\ell$ is still not known
in this low-carrier-density compound, the presence of clear
Andreev-reflection features, the general shape of the conductance
curves, the absence of dips \cite{Sheet_Dips} and the coincidence
between the critical temperature of the junction and the bulk T$_c$
indicate that the spectroscopic requirements are satisfied in our measurements.\\

%%%%%%%%%%%%%%%%%%%%%%%%%%%%%%%%%%%%%%%%%%%%%%%%%%%%%%%%%%%%%%%%%%%
\vspace{-2mm}
%%%%%%%%%%%%%%%%%%%%%%%%%%%%%%%%%%%%%%%%%%%%%%%%%%%%%%%%%%%%%%%%%%%%%

The differential conductance curves of our point contacts were
obtained by taking the numerical derivative of the I-V
characteristics. In order to compare the results with a theoretical
model, the measured conductance curves of each junction had first to
be normalized, i.e. divided by the relevant normal-state
conductance. A fit was then performed by means of a two-band
modified \cite{Plecenik} Blonder-Tinkham-Klapwijk model \cite{BTK}
generalized to take into account the angular distribution of the
current injection at the N/S interface \cite{Kashiwaya}. According
to this model the normalized conductance is the weighed sum of the
conductances of the two bands, $G=w_1G_{1}^{BTK}+(1-w_1)G_{2}^{BTK}$
where $w_1$ is the weight of band $1$. Each conductance depends on 3
parameters: the gap value, $\Delta$, a broadening parameter,
$\Gamma$ \cite{Plecenik}, and the barrier parameter $Z$ that
accounts for both the height of the potential barrier and the
mismatch of the Fermi velocities at the N/S interface
\cite{BTK,Kashiwaya}. Although the model contains 7 fitting
parameters (3 for each band plus $w_1$), they are not totally free
in the sense that, if one fits the whole temperature dependence of a
conductance curve, the two barrier  parameters Z$_1$ and Z$_2$ as
well as the weight $w_1$ should be kept constant with increasing
temperature, while the broadening parameters $\Gamma_1$ and
$\Gamma_2$ should remain almost constant, or increase at
T$>$T$_c/2$. These requests automatically restrict the variability
of these parameters.

%\section{Results and discussion}
Figure \ref{Fig1} shows the raw conductance curves of a 28-$\Omega$
contact, measured from 4.2 K up to about 180 K. The critical
temperature of the junction is $\sim$ 52 K. The low-temperature
curve (upper thick line) shows clear Andreev-reflection features
such as two peaks at $\sim \pm$6 mV, certainly related to a
superconducting gap \cite{Sm_Nature}, plus two pronounced shoulders
at higher bias voltages that, in a way very similar to what
previously observed in MgB$_2$ \cite{nostroPRL}, can indicate the
presence of a second, larger gap (this indication will be further
substantiated in the following, by fitting the conductance curves up
to T$_c$ and observing the temperature evolution of the gaps). In
all our contacts, the normal-state conductance curve  measured at
T$_c$ (middle thick line in Fig. \ref{Fig1}) features a hump at zero
bias that gradually decreases with increasing temperature until it
completely disappears (lower thick line) around the N\'{e}el
temperature of the parent compound, T$_N$ $\sim 140$ K
\cite{Drew_PhDiagr_Sm}. This might suggest a magnetic origin of this
hump. A similar downward curvature of the normal-state spectrum was
also found in recent PCAR \cite{Szabo_BaKFeAs} and ARPES
\cite{Ding_ARPES_BaKFeAs} measurements in
Ba$_{0.6}$K$_{0.4}$Fe$_2$As$_2$. Finally, the conductance is
asymmetric (higher at negative bias, i.e. when electrons are
injected into the superconductor) at all temperatures, even above
T$_c$. This asymmetry, clearly visible in Fig.\ref{Fig1}, was also
observed in other PCAR measurements \cite{LaOFFeAS,Szabo_BaKFeAs}
and is most probably related to the fast decrease of the density of
states at the Fermi level \cite{Singh}. Note that the asymmetry
persists even when the zero-bias hump disappears.

\begin{figure}[t]
\begin{center}
\includegraphics[keepaspectratio, width=0.8 \columnwidth]{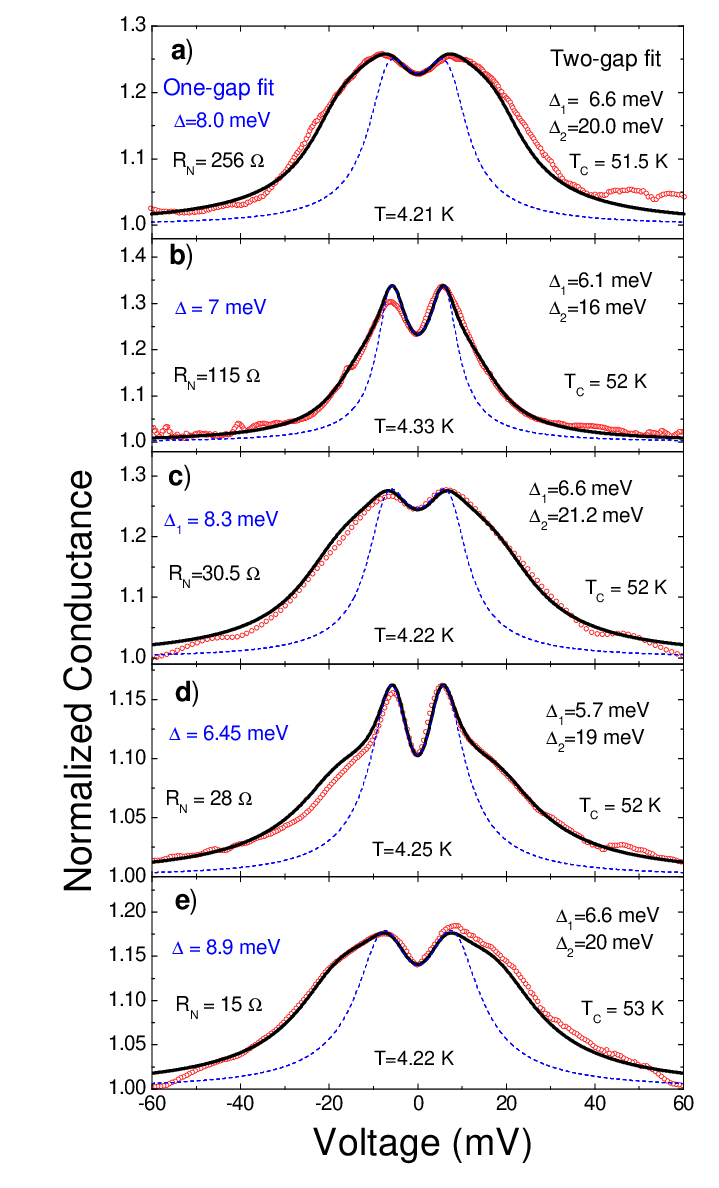}
\end{center}
\vspace{-5mm} \caption{\small{(color online) (a) to (e): normalized
conductance curves (symbols) at 4.2 K in SmFeAsO$_{0.8}$F$_{0.2}$.
The contact resistance, R$_N$, decreases from top to bottom. The
curves are reported together with their relevant one-gap (dash
lines) and two-gap (solid lines) BTK fits. The gap values are
indicated on the left (single-gap fit) and
on the right (two-gap fit).}} \label{Fig2} %\vspace{-5mm}
\end{figure}

Figure \ref{Fig2} shows several examples of low-temperature
normalized conductance curves (symbols). Since the upper critical
field, H$_{c2}$, is certainly very high \cite{Senatore_Hc2}, the
normal state conductance at T$ < $T$_c$ is not experimentally
accessible. The normalization was therefore performed by using the
normal-state conductance curve measured at the critical temperature
of the junction. The contact resistance R$_N$, indicated in the
labels of each panel, decreases from top to bottom. In all cases the
one-gap BTK fit (dash lines) is compared to the two-gap one (solid
lines). It is clear that the quality of the fit performed with the
one-gap model is quite poor and can account only for a small central
portion of the curve (as in Ref.\cite{Sm_Nature}). The resulting
values of the gap range between 6.45 and 8.9 meV, corresponding to
$2\Delta(0)/k_BT_c$= 2.9-3.9. The two-gap fit can instead reproduce
the measured conductances remarkably well and gives values of
5.7-6.6 meV for the small gap, and of 16-21 meV for the large one.
In most cases, the amplitude of the Andreev signal (height of the
curves) is of the order of 20-30 \% (similar to what observed in
Refs. \cite{Sm_Nature,nostroPRL,CaC6}) and gives clear and
unambiguous measurements; the ratio $\Gamma(0)/\Delta(0)$ was
usually $\sim 0.5-0.6$ or smaller; $w_1$ was $0.55 \pm 0.10$
depending on the junction. The normalized conductances often show a
residual asymmetry between the left and the right part which mainly
affects the determination of the larger gap; this asymmetry might be
due to the aforementioned unconventional shape of the background
which is itself asymmetric and changes with temperature. When
possible, a fit of both sides of the normalized curves was done in
order to determine the spread of $\Delta_2$ values arising from this
asymmetry.

%%%%%%%%%%%%%%%%%%%%%%%%%%%%%%%%%%%%%%%%%%%%%%%%%%%%%%%%%%%%%%%%%%%
%\vspace{-5mm}
%%%%%%%%%%%%%%%%%%%%%%%%%%%%%%%%%%%%%%%%%%%%%%%%%%%%%%%%%%%%%%%%%%%%%

Several point-contact measurements on Fe-As-based superconductors
reported a dependence of the conductance curves on the contact
resistance and, therefore, on the pressure applied by the tip with
the consequent appearance of additional structures, mainly zero-bias
conductance peaks (ZBCP), not related to the superconducting gaps
\cite{Sm_Nature,Yates_PCS_Nd}. As clearly shown in Fig.\ref{Fig2},
this problem is completely overcome in our point contacts, where no
pressure is applied to the sample. The general shape of our curves
is rather independent on the contact resistance (as expected for
ballistic junctions) and the extracted gap values are in good
agreement with each other even for very different contact
resistances (see, for example, the curves in panel (a) and (e) of
Fig. 2). The systematic absence of zero-bias peaks in our curves
completely rules out the $d$-wave symmetry for the gaps. As a matter
of fact it can be easily shown that, in the case of a $d$-wave gap,
all the conductance curves calculated within the generalized BTK
model \cite{Kashiwaya} using $\Gamma(0)/\Delta(0)=0.5-0.6$ and Z
values similar to the experimental ones (Z = 0.3-0.4), show clear
ZBCP at 4.2 K whenever the direction of injection of the current
forms an angle $\beta>15^{\circ}$ with respect to the anti-nodal
direction \cite{Stefanakis}. If the gaps had $d$-wave symmetry, the
experimental absence of ZBCP would mean that the current was always
injected within $15 ^{ \circ } $ of the anti-nodal direction, which
is absolutely unreasonable in the case of polycrystalline samples.
%
%%%%%%%%%%%%%%%%%%%%%%%%%%%%%%%%%%%%%%%%%%%%%%%%%%%%%%%%%%%%%%%%%%%%
%\vspace{-5mm}
\begin{figure}[t]
\begin{center}
\includegraphics[keepaspectratio, width=0.9 \columnwidth]{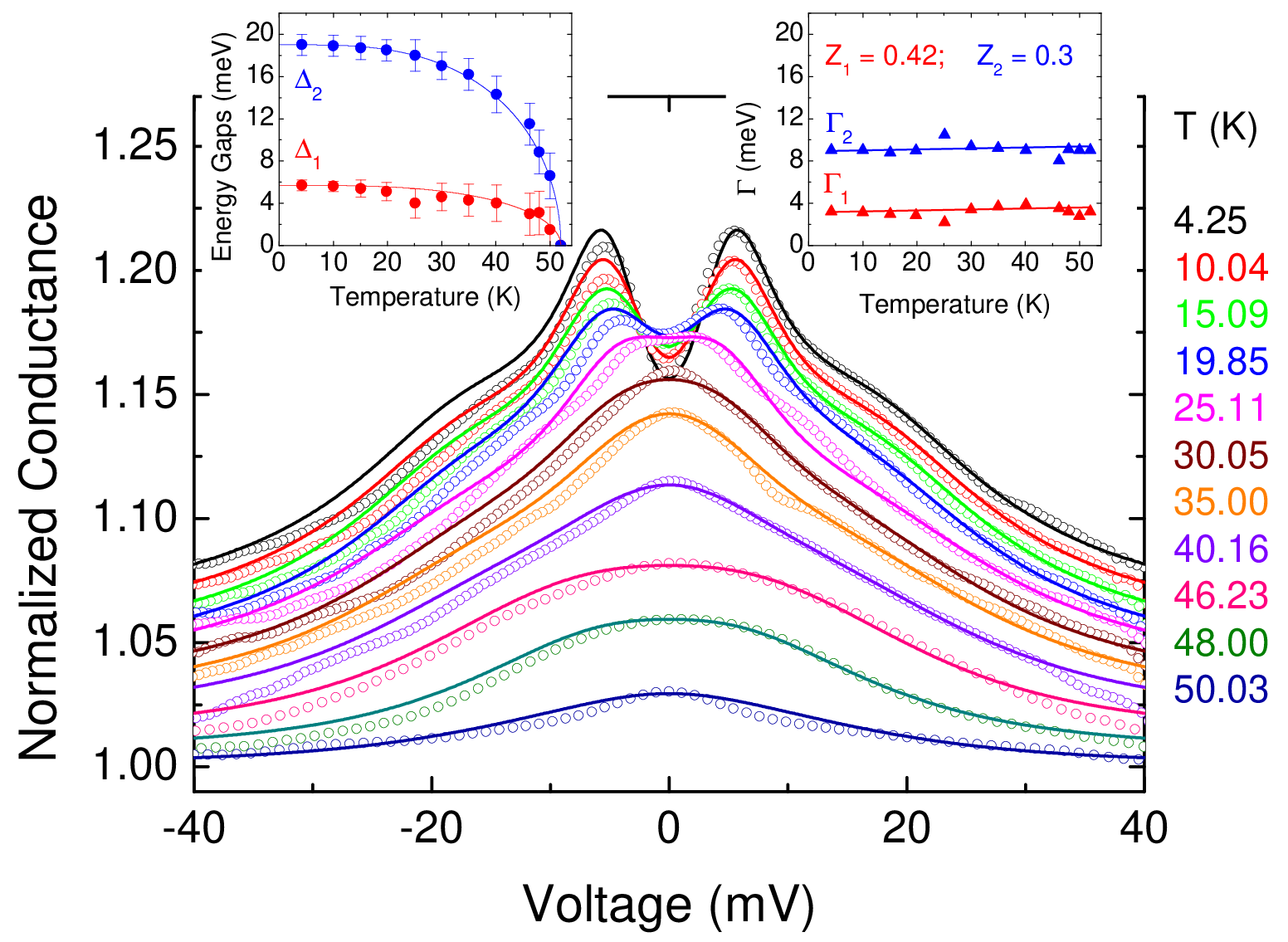}
\end{center}
\vspace{-5mm} %
\caption{\small{(color online) Temperature dependence of the
normalized conductance curves (symbols) together with their relevant
two-gap BTK fits (lines). The curves are vertically shifted for
clarity. The raw data are those shown in Fig. 1 and the insets show
the temperature dependency of the gaps ($\Delta_1$ and $\Delta_2$)
and of the broadening parameters ($\Gamma_1$ and $\Gamma_2$) given
by the two-gap fit.}} \label{Fig3} \vspace{-5mm}
\end{figure}
%%%%%%%%%%%%%%%%%%%%%%%%%%%%%%%%%%%%%%%%%%%%%%%%%%%%%%%%%%%%%%%%%%%%%

In order to obtain the behavior of the two gaps as a function of
temperature, a fit of the temperature dependence of the normalized
conductance curves was performed with the two-band BTK model. A
typical result is shown in Fig. \ref{Fig3} where the experimental
data (symbols) are compared to the relevant BTK curves (lines) that
best fit the positive-bias side. The barrier parameters Z$_1$ and
Z$_2$ used for the fit are indicated in the label of the right
inset; a small decrease (of the order of 20\% of these values) had
to be allowed to obtain a good fit of the whole temperature
dependence -- but this might simply mean that the normal-state
conductance at low temperature is more peaked that that at T$_c$.
The values of the gaps $\Delta_1$ and $\Delta_2$ and of the
broadening parameters $\Gamma_1$ and $\Gamma_2$ are reported in the
insets. Note that the temperature dependence of the gaps is
impressively BCS-like, with gap ratios $2\Delta_1 (0)/ k_B T_c=2.55$
and $2\Delta_2 (0) / k_B T_c=8.5$. Here, error bars indicate the
uncertainty on the gaps due to the fit \emph{and} the experimental
resolution of the technique, $k_B T$.

%%%%%%%%%%%%%%%%%%%%%%%%%%%%%%%%%%%%%%%%%%%%%%%%%%%%%%%%%%%%%%%%%%%%
\begin{figure}[t]
\begin{center}
\includegraphics[keepaspectratio, width=0.9 \columnwidth]{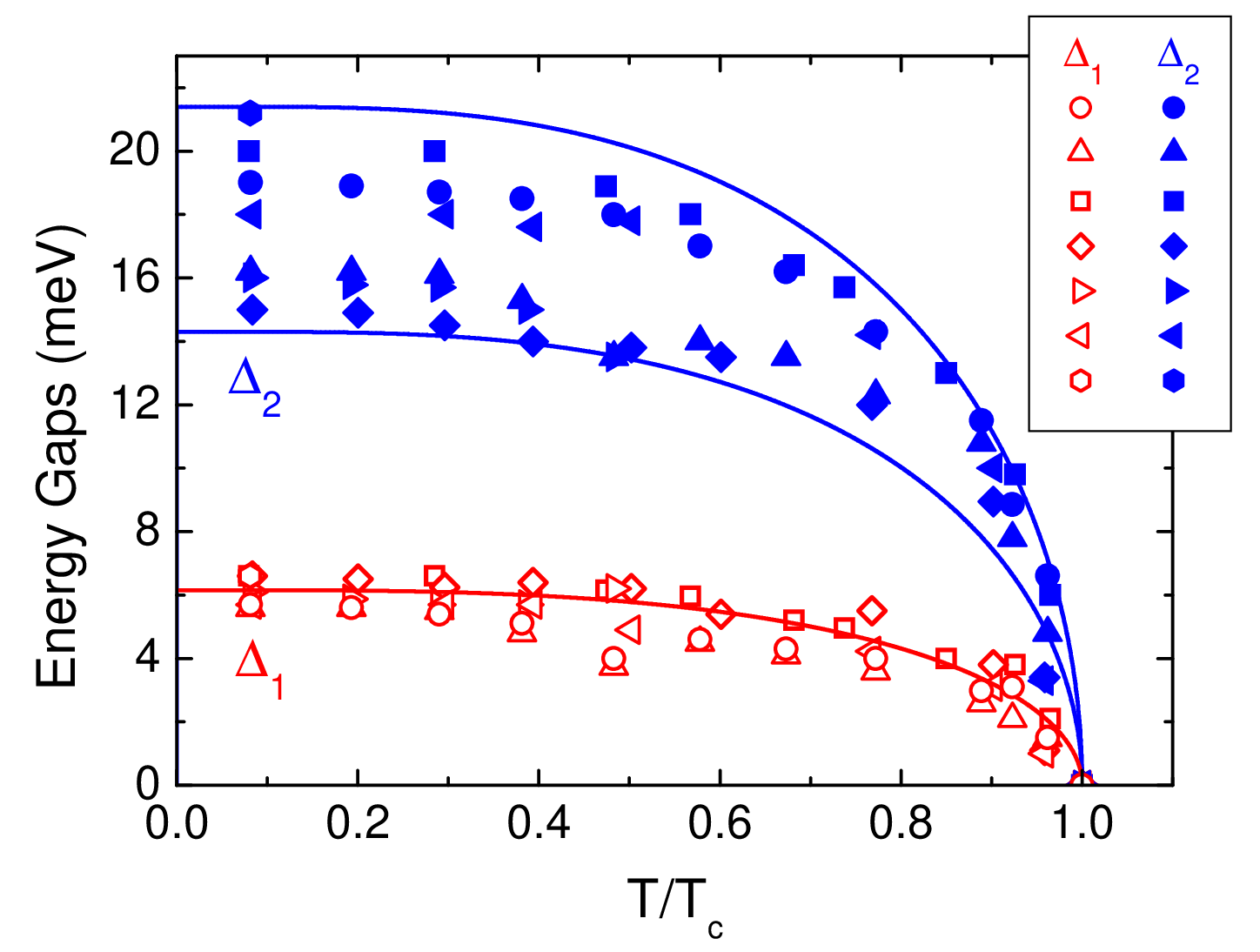}
\end{center}
\vspace{-5mm} %
\caption{\small{(color online) Temperature dependence of the gaps
$\Delta_1$ (open symbols) and $\Delta_2$ (full symbols) as extracted
from the two-gap fit of the conductance curves. Lines are BCS-like
curves. Details are in the text.}}
\label{Fig4} %\vspace{-5mm}
\end{figure}
%%%%%%%%%%%%%%%%%%%%%%%%%%%%%%%%%%%%%%%%%%%%%%%%%%%%%%%%%%%%%%%%%%%%%

The temperature dependence of the gaps obtained from the two-gap BTK
fit of the curves shown in figure \ref{Fig3} and from various other
measurements is reported in figure \ref{Fig4}. The reproducibility
of the small gap (open symbols) is noticeable: $\Delta_1$ is close
to 6 meV at low temperature and follows a BCS-like trend up to the
critical temperature which was always between 50 and 53 K (the
normalized temperature T/T$_c$ is used in Fig.\ref{Fig4} for
homogeneity). As far as $\Delta_2$ is concerned, its behaviour in
each set of data is nicely BCS, but the spread of values between
different data sets is rather wide: all the data fall in a region
bounded by two BCS-like curves with gap ratios $2\Delta_2 (0) /k_B
T_c\simeq 7$ and 9, respectively. Incidentally, this region is
centered on a value $\overline{\Delta_2}\simeq 18$ meV that is close
to that determined with infrared ellipsometry
\cite{Dubroka_LargeGap}, $\Delta \sim 18.5$ meV. The spread of
$\Delta_2$ values shown in Fig.\ref{Fig4} cannot be ascribed to our
PCAR technique, since in MgB$_2$, for example, both the gaps were
determined in exactly the same way with a great accuracy ($\pm 0.5$
meV at most, at 4.2 K). It is instead a property of this particular
compound; it mainly comes from the residual asymmetry of the
normalized curves (when the negative/positive bias side of the same
curve is fitted). Even though, as explained above, this asymmetry is
likely to be due to the shape of the normal state (which possibly
becomes more and more asymmetric on decreasing temperature), we
cannot exclude other possibilities, i.e. that it is an intrinsic
feature of the superconducting state in this material. Finally,
notice that at low temperature the ratio of the gaps,
$\Delta_2/\Delta_1$ is $\sim 3$. Similar values have also been
reported for other pnictide superconductors
\cite{Szabo_BaKFeAs,LaOFFeAS,Kawasaki_NQR_LaOFFeAS} and might
indicate that the proposed interband-only, extended s$\pm$-wave
model \cite{Mazin_spm} cannot fully explain the experimental data
since a higher intraband coupling of non-magnetic origin should also
be taken into account \cite{Dolgov_BCSvsEliash}.

%\section{Conclusions}

In summary, we performed pressure-less PCAR measurements in
SmFeAsO$_{0.8}$F$_{0.2}$ with T$_c$ up to 53 K. The experimental
point-contact conductances show peaks around $\pm$ 6 mV and clear
shoulders at higher bias voltages, $\pm$ 16 - 20 mV, indicating the
presence of two superconducting gaps. The BTK fit of the
low-temperature normalized conductances fully supports this
observation: the single-band fit gives gap values in good agreement
with those reported in literature \cite{Sm_Nature} but can only
reproduce a small central portion of the conductance curve. The
two-gap fit, instead, accounts remarkably well for the shape of the
whole experimental curve. Furthermore, the fit of the temperature
dependence of the normalized conductance curves gives a BCS-like
dependence for both the gaps, which close at the same critical
temperature, directly revealing that multi-gap nodeless
superconductivity is a fundamental property of this oxypnictide
superconductor. Averaging over various contacts, we obtained
$\Delta_1(0) = 6.15 \pm 0.45$ meV and $\Delta_2(0) = 18 \pm 3$ meV
with ratios $2\Delta_1 (0)/k_B T_ c = 2.5 - 3$ and $2\Delta_2
(0)/k_B T_c = 7 - 9$, respectively. This result, together with other
experimental findings in pnictide superconductors
\cite{Hunte,Ding_ARPES_BaKFeAs,Szabo_BaKFeAs,LaOFFeAS,Kawasaki_NQR_LaOFFeAS,Matano_Pr_NMR,Samuely_Nd_PCAR,Pan_Nd_STM,Wang_PCS},
points therefore towards a common multi-gap scenario to most (if not
all) Fe-As-based superconductors.

We thank G.A. Ummarino, I.I. Mazin, P. Szab\'{o}, L.F. Cohen and Y.
Fasano for useful discussions. V.A.S. acknowledges support by the
Russian Foundation for Basic Research Project No. 09-02-00205. This
work was done within the PRIN project No. 2006021741.

\end{document}